\documentclass[twoside,12pt]{article}
\usepackage{epsf,epsfig,amssymb,amsmath,graphicx,float}
\usepackage{color}
\usepackage{hyperref}

\usepackage{indentfirst}
\begin{document}
\renewcommand{\thefootnote}{\fnsymbol{footnote}}

\begin{titlepage}

\begin{center}

\vspace{1cm}

{\Large {\bf Constraints on non-standard cosmological models from Planck data}}

\vspace{1cm}

{\bf Hoernisa Iminniyaz$^a$, Alimasi Aisha$^b$, Fangyu Liu$^a$\footnote{Corresponding 
author, l.fangyu@qq.com}}

\vskip 0.15in
{\it
$^a$
{School of Physics Science and Technology, Xinjiang University, \\
Urumqi 830017, China} \\
$^b$
{College of Electronic and Optical Engineering $\&$ College of Flexible
  Electronics (Future Technology), Nanjing University of Posts and
  Telecommunications, \\
Nanjing 210046, China} \\

}

\abstract{
We review the relic density of dark matter in the non-standard cosmological
scenarios which includes kination models, brane world cosmology and shear 
dominated universe. Then we use the Planck data to find constraints on the
parameter spaces as dark matter cross sections and the five dimentional Planck mass for
brane cosmology, enhancement factor for kination model and the inverse-scaled 
shear temperature for shear dominated universe.  }
\end{center}
\end{titlepage}
\setcounter{footnote}{0}

\section{Introduction}

Recent data by Planck team provided that the cold Dark Matter (DM) relic 
density is   
\begin{equation}\label{eq:omega1}
      \Omega_{\rm DM} h^2 = 0.120 \pm 0.001, 
\end{equation}
where $ h = 0.710 \pm 0.025 $ is the scaled Hubble constant in units of 
$100$ km s$^{-1}$ Mpc$^{-1}$ \cite{Planck:2018vyg}.
This value has significant importance for the particle physics models which 
possess DM candidates. Theoretically predicted DM relic density
should fall in the range of Eq.(\ref{eq:omega1}). The parameter spaces of the
existing models are constrained by this requirement.  

The long-lived or stable weakly interacting massive particles (WIMPs) in weak-scale mass are proposed as good DM candidates. WIMPs are 
considered to be in thermal equilibrium with the standard model particles in 
the early universe and they decoupled from thermal equilibrium during the 
radiation dominated era. Based on this assumptions, the precise analytic
approximate solutions are derived for the relic abundance of DM 
\cite{standard-cos}. We found the relic abundance of DM depends not only on the
annihilation cross sections, but also the cosmological
parameters. Particularly the Hubble expansion rate plays important role to 
determine the relic density of DM. 

The Hubble expansion rate is fixed by the Friedmann equation of general
relativity in the standard cosmological scenario. In this scenario, according 
to the assumption, the DM particles decouple from the 
thermal equilibrium at the freeze out point where the interaction rate of
the DM particles falls below the Hubble expansion rate of the universe. The
decoupling point of WIMPs
 is usually considered to be earlier than Big Bang
Nucleosynthesis (BBN). Unfortunately, we do not have
observational cosmic information or evidence that the universe must be radiation
dominated prior to BBN. Thus, our attention is drawn to non-standard
cosmological models which predict the expansion rate of the universe is
different from the standard case before BBN, and returns to the standard case at the beginning of BBN. 
 There are several non-standard
cosmological models in which the Hubble expansion rate is increased, for
example, kination model, brane world cosmology, 
shear dominated universe and etc
\cite{Salati:2002md,Randall:1999vf,Langlois:2002bb,Okada:2004nc,AbouElDahab:2006glf,Bianchi1,Bianchi2,Bianchi,Kamionkowski:1990ni,Schelke:2006eg,Catena:2009tm}. In these models, the relic density of DM is increased due to the
modification of the Hubble expansion rate.
In kination models, the energy density of the universe 
is dominated by the kinetic energy of the scalar field. The impact of kination
period on the relic abundance of DM was discussed in \cite{Salati:2002md}. 
Brane world
cosmology is another non-standard cosmological scenario in which the ordinary
matter is supposed to be confined onto a three-dimensional subspace which embedded in a 5 dimensional spacetime. The extra 
dimension has impact on the relic density of DM in brane world cosmology 
\cite{Okada:2004nc,AbouElDahab:2006glf}. In shear dominated universe, the 
expansion of the universe is not
isotropic. The anisotropy also leaves its imprint on the relic density of DM
\cite{Kamionkowski:1990ni}. The characteristic of all the models we discussed 
above is that the
Hubble expansion rate is enhanced and there are increased relic density of DM
in the end.           

Since these non-standard cosmic eras leads to the deviation in DM abundance predicted by standard cosmology, for obtaining the observed DM abundance, it is necessary to constrain the parameters of non-standard models. That is the purpose of this article. We have precise measurement of the DM relic density as 
in Eq.({\ref{eq:omega1}}). Whenever the WIMP annihilation cross section is 
determined, we can constrain the parameter spaces in those models like
enhancement factor in kination model, 5 dimensional Planck mass
in brane world cosmology and the inverse-scaled shear temperature in shear 
dominated universe.

The paper is arranged as following. In section 2, we review the evolution of the cosmic expansion rate in non-standard cosmological scenarios including kination
model, brane world cosmology and the shear dominated universe. In section 3,
we review the DM relic density calculation in non-standard cosmological
models. In section 4, we use the Planck data to find constraints on the 
enhancement factor for
kination, 5 dimensional Planck mass in brane world cosmology, and the
inverse-scaled shear temperature in shear dominated universe.
 The conclusion and summaries are in the last section.

\section{Hubble expansion rate in non-standard cosmological scenarios}
In this section, we briefly review the Hubble expansion rate in non-standard
cosmological scenarios. First we start with the kination model. The period of
early universe in which the kinetic energy of a scalar field dominated is
called ``kination''. Its action is $I_{\Phi}=-\int
d^{4}x\sqrt{-Detg}(\frac{1}{2}g^{ \mu\nu}\frac{\partial\Phi}{\partial x^{
    \mu}}\frac{\partial\Phi}{\partial x^{\nu}}+V)$. Applying the 
action
principle, one can obtain the field equation. Then calculating the field equation, the energy density of the scalar field can be obtained, i.e. $\rho_{\Phi}\propto R^{-6}$ when the kinetic energy of the scalar field is much larger than the scalar potential (For details of the calculations, one can see the Refs.\cite{Joyce:1996cp,Spokoiny:1993kt}). Here, $R$ is the scale factor of the universe.
Therefore, during kination period, 
$H_{\rm k}^2 \propto \rho_{\rm tot} \simeq \rho_{\Phi} \propto R^{-6}$. The relation
between the expansion rate $H_{\rm k}$ in kination period and the expansion rate
$H_{\rm std}$ in standard cosmology is given by    
\begin{equation} \label{eq:Hk}
    H_{\rm k} = H_{\rm std} \sqrt{ 1 + \frac{\rho_{\Phi}}{\rho_r}},   
\end{equation}    
where 

\begin{equation} \label{eq:Hkr}
    \frac{\rho_{\Phi}}{\rho_r} \simeq \eta \left( \frac{T}{T_0} \right)^2=\eta\left (\frac{x_0}{x}\right )^2,
\end{equation}
here $T_0$ is the reference temperature, and $x = m/T$, $m$ is the mass of DM particle. $\rho_{\rm r} = \pi^2/30\, g_* T^4$ is the energy density of radiation, $H_{\rm std}$ is 
\begin{equation}
      H_{\rm std} = \pi \sqrt{\frac{g_*}{90}} \frac{T^2}{M_{\rm Pl}},
\end{equation}
where $M_{\rm Pl} =2.4 \times 10^{18}$GeV is the reduced Planck mass and $g_*$ is the effective number of relativistic degrees of freedom. 

Next, we review the Hubble expansion 
rate in brane world cosmological model. In this model, the standard model
particles are supposed to be confined onto a three-dimensional subspace, it is
called brane that is embedded in a higher dimensional spacetime.
In the Gaussian Normal coordinate system, its line-element can be
  written as 
$ds^2=-h^2(t,y)dt^2+R^2(t,y)\gamma_{ij}dx^idx^j+f^2(t,y)dy^2$. $R(t,0)$ is the cosmic scale factor , when the brane is located at $y=0$.
Calculating the Einstein field equation and using the junction conditions, the 
evolution of the cosmic scale factor can be obtained 
(For details of the calculations, one can see the
Refs.\cite{Binetruy:1999ut,Binetruy:1999hy}). The expansion rate in this model
is quite different
from the expansion rate in the standard cosmology. The brane energy density
contributes quadratically to $H_{\rm b}^2$ in this model, while the ordinary energy
density contributes linearly to $H_{\rm std}^2$ in the standard model.
The Hubble expansion rate is given as 
\begin{equation} \label{eq:Hb}
    H_{\rm b} = H_{\rm std} \sqrt{ 1 + \frac{\rho_{\rm r}}{2 \lambda} }\,,
\end{equation}
where $\lambda = 6  M_5^6/M^2_{\rm Pl}$. It is simplified as
\begin{equation} \label{eq:Hbsimp}
    H_{\rm b} = H_{\rm std} \sqrt{ 1 + \frac{k_{\rm b}}{x^4} }\,.
\end{equation}
Here $k_{\rm b} = \pi^2 g_* m^4 M^2_{\rm Pl}/(360 M_5^6) $, where $M_5$ is the 5
dimensional Planck mass. 

The third modified cosmological model is the Bianchi type I model which
assumes the universe is homogenous but anisotropic before BBN. Its line-element is 
$ds^2=-dt^2+R_1^2dx^2+R_2^2dy^2+R_3^2dz^2$, where $R_{\rm i}$ is the scale factor
associated with $x,\,y,\,z$. Calculating the Einstein field equation and
taking the energy-momentum tensor form as a perfect fluid, the evolution of
scale factor $R_{\rm i}$ can be obtained (For details of the calculations, one can
see the
Refs.\cite{Hertzberg:2024uqy,Gron:2024vmf,Kamionkowski:1990ni}). We
mention that in this model, the Liouville operator $L[f]=p^\mu\partial_\mu f-{\Gamma^\mu}_{\sigma\nu}p^\sigma p^\nu\frac{\partial f}{\partial p^\mu}$ which is the part of the differential form of the Boltzmann equation differs from the case in the standard cosmology, but the form of Boltzmann equation which describes the
time evolution of the number density of particles is the same (just substituting $H_{\rm s}=(H_1+H_2+H_3)/3$ for $H_{\rm std}$, where $H_{\rm i}\equiv (dR_{\rm i}/dt)/R_{\rm i}$. For details of the calculations, one can see the Refs.\cite{Kamionkowski:1990ni,reviewBE,Enomoto:2023cun}). In this model, the shear-energy density is propotional to the amount of anisotropic expansion. When
$T \gg T_{\rm e}$, the universe is shear dominated at which point 
$H_{\rm s} \propto \bar{R}^{-3} $ and $\bar{R} \propto t^{1/3}$, where $T_{\rm e}$ is the reference temperature for $\rho_{\rm r}=\rho_{\rm s}$ and $\bar{R}\equiv(R_1R_2R_3)^{1/3}$;
when $T \ll T_{\rm e}$, the universe is
radiation dominated, here the expansion rate $ H_{\rm s} \propto \bar{R}^{-2}$ and
$\bar{R} \propto t^{1/2}$.
In this scenario, the Hubble expansion rate is derived as 
\begin{equation} \label{eq:Hs}
     H_{\rm s} = H_{\rm std}\sqrt{1 + \frac{\rho_{\rm s}}{\rho_{\rm r}}} 
\end{equation}
The shear-energy density is expressed in terms of the radiation
energy density as 
\begin{equation} \label{eq:rho}
     \rho_{\rm s} = \rho_{\rm r} \left[ \frac{g_* T^2}{ g_*^e T^2_{\rm e}}  \right],
\end{equation} 
where $g_*^e $ is the value of $g_*$ at $T_{\rm e}$.
Eq.(\ref{eq:Hs}) is simplified as
\begin{equation}
     H_{\rm s} = \frac{\pi m^2}{M_{\rm Pl} x^2} 
     \sqrt{\frac{g_*}{90} \left( 1 + \frac{x_{\rm e}^2}{x^2} \right)}
\end{equation} 
We can write the Hubble expansion rate in modified cosmological scenarios
as 
\begin{equation}\label{eq:ModH}
     H = A(x)H_{\rm std},
\end{equation}
here $A(x) = \sqrt{ 1 + \eta (x_0/x)^2}\,\,\,, \sqrt{ 1 +  k_{\rm b}/x^4}\,\,\,, 
\sqrt{ 1 +  x^2_{\rm e}/x^2}$ for kination model, brane world model and Bianchi 
type I models respectively. 
%
 %
%
%
%

\section{Relic density of DM in non-standard cosmological scenarios}
The relic density of DM in non-standard cosmological
scenarios was discussed in
\cite{Salati:2002md,Okada:2004nc,AbouElDahab:2006glf,Kamionkowski:1990ni,Schelke:2006eg,Catena:2009tm}. The
relic density of DM is obtained 
by solving the Boltzmann equation. 
\begin{eqnarray} \label{eq:boltzmann_n}
\frac{{\rm d} n_{\chi}}{{\rm d}t} + 3 H n_{\chi} =  
- \langle \sigma v\rangle (n^2_{\chi}  - 
n^2_{\chi,{\rm eq}} )\,,
\end{eqnarray}
The equilibrium number density is given as 
\begin{equation} 
n_{\chi,{\rm eq}} = g_\chi \left(\frac{m T}{2 \pi}\right)^{3/2}\,,
\end{equation}
where $g_{\chi}$ is the number
of intrinsic degrees of freedom of the particle, $n_{\chi}$ is the number density of DM particle. 
In nonrelativistic cases, the thermal averaged annihilation cross section is
\begin{equation} \label{eq:cross}
     \langle  \sigma v \rangle  \approx  a + 
                           \frac{6 b}{x}.
\end{equation}
Here $a$ is the $s-$wave contribution when $v \to 0$ and $b$ is the $p-$wave
contribution when $s-$wave is suppressed. 
In the standard frame, the DM particles are assumed to be in thermal equilibrium
with the standard model particles in the early universe when the interaction rate satisfies $\Gamma> H$. DM
particles decouple from the equilibrium state when  $\Gamma < H $. From that point, the DM relic density almost
becomes constant. The temperature at which the DM decouple from equilibrium is 
called freeze out temperature $T_F$. 
Boltzmann equation (\ref{eq:boltzmann_n}) can be
written in terms of the dimensionless quantities $Y_{\chi} = n_{\chi}/s$, here the entropy density $s$ is given as $s = 2 \pi^2 g_{*s}/45\, T^3$
with $g_{*s}$ being the entropic degrees of freedom. Assuming entropy is conserved, 
Eq.(\ref{eq:boltzmann_n}) is rewritten as
\begin{equation} \label{eq:boltzmann_Y}
\frac{{\rm d} Y_{\chi}}{{\rm d}x} =
      - \frac{\Lambda \langle \sigma v \rangle}{A(x) x^2}\,
     (Y^2_{\chi} - Y^2_{\chi, {\rm eq}}   )\,,
\end{equation}
where $\Lambda = 1.32\,m M_{\rm Pl}\, \sqrt{g_*}\,$,
 $g_*\simeq g_{*s}$ and $dg_{*s}/dx\simeq 0$. 
The final relic abundance for DM particle is derived from 
Eq.(\ref{eq:boltzmann_Y}) as
\begin{equation} \label{eq:barY_cross}
Y_{\chi}(x_\infty) =   \frac{1}{1.32 m M_{\rm Pl}\sqrt{g_*} \int^{\infty}_{x_F}
  {\rm d}x {\langle \sigma v \rangle}/(A(x)x^2)}  
\end{equation}
The detailed analysis is given in \cite{standard-cos}. We are not going to 
discuss it here. The freeze out temperature $x_F$ is determined by
assuming that the freeze out occurred at the point when 
the deviation of the relic abundance $Y_{\chi}(x_F)$ and 
$Y_{\chi,{\rm eq}}(x_F)$ is of the same order as the 
equilibrium value
such as $Y_{\chi}(x_F) = (1+\xi) Y_{\chi,{\rm eq}}(x_F) $.
Here $\xi$ is a constant and usually we take 
$\xi=  \sqrt{2} -1 $ \cite{standard-cos}. 

The final DM relic density is expressed as  
\begin{equation} \label{eq:omega}
 \Omega_{\rm DM}  h^2  =  2.76 \times 10^8\, m Y_{\chi}(x_\infty)\,,                
\end{equation}
where $\Omega_{\chi} = \rho_{\chi}/\rho_{\rm c}$ with 
$\rho_{\chi}=n_{\chi} m = s_0 Y_{\chi} m $ and 
$\rho_{\rm c} = 3 H^2_0 M^2_{\rm Pl}$, here $s_0 \simeq 2900$ cm$^{-3}$ is the 
present entropy density, and $H_0$ is the Hubble constant.

\section{Constraints }
\begin{figure}[t!]
  \begin{center}
    \hspace*{-0.5cm} \includegraphics*[width=8cm]{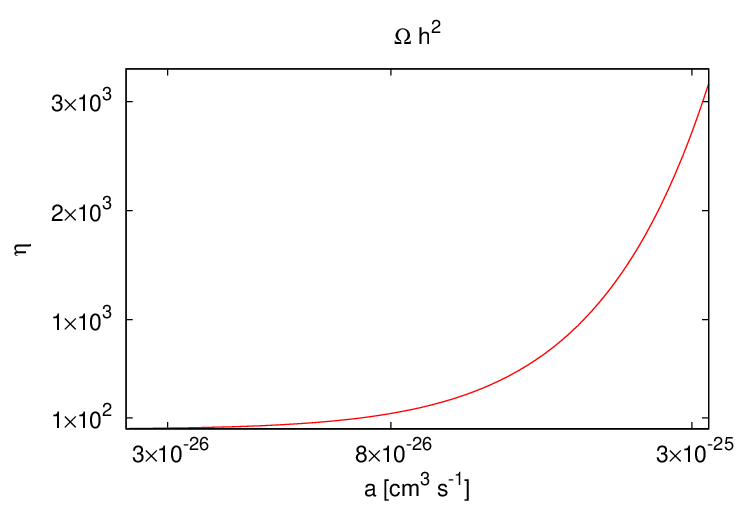}
    \put(-115,-12){(a)}
    \hspace*{-0.5cm} \includegraphics*[width=8cm]{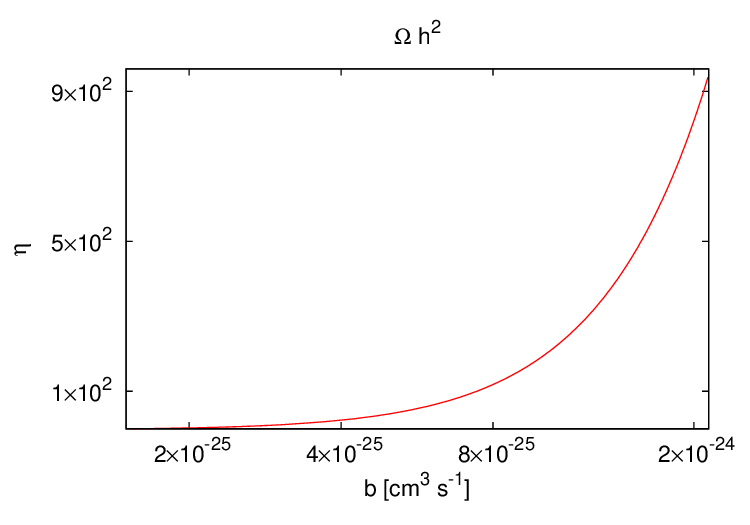}
    \put(-115,-12){(b)}
    \caption{\label{fig:a} 
    \footnotesize Contour plots of the $s-$wave ($p-$wave) annihilation cross 
    section $a$ ($b$) and 
    the enhancement factor $\eta$ when 
    $\Omega_{\rm DM} h^2 = 0.120$ \cite{Planck:2018vyg}. Here  
    $m_{\chi} = 100$ GeV,
    $g_{\chi} = 2$, $g_* = 90$, $x_0 = 20$. }
     \end{center}
\end{figure}

Following, we use the Planck data Eq.(\ref{eq:omega1}) to constrain 
the parameter spaces of the modified cosmological scenarios. 
In Fig.\ref{fig:a}, the relation between the enhancement factor $\eta$ of
kination model and the cross section of DM is shown when $\Omega_{\rm DM} h^2 = 0.120$. 
Panel $(a)$ is for $s-$wave annihilation cross section and panel $(b)$ is for
$p-$wave case. In panel $(a)$ in Fig.\ref{fig:a}, when the cross section takes 
the values from $2.08 \times 10^{-26} {\rm cm^3} {\rm s}^{-1}$  to 
$3.26 \times10^{-25}{\rm cm^3} {\rm s}^{-1}$, 
the enhancement factor $\eta$ varies from $5.00 \times 10^{-2}$ to 
$3.24 \times 10^3$.
The enhanced Hubble rate let the DM particles decouple
from the thermal equilibrium earlier in kination model. Then there
is increased relic density. The size of the increase depends on the
enhancement factor. The detailed analysis was done in \cite{Salati:2002md}. 
When the enhancement factor
is small, the increase is not notable, then the related cross section is not
large. On the contrary, for larger $\eta$, there is significant increase of
the relic density. In order to obtain the observed relic density of DM, we
need larger cross section. This is the reason why for larger $\eta$, there is
larger $a$. The same rule is applied for the case of $p-$wave annihilation. 
For $p-$wave annihilation cross sections ranged from  
$1.54 \times 10^{-25} {\rm cm^3} {\rm s}^{-1}$ to 
$2.13 \times 10^{-24} {\rm cm^3} {\rm s}^{-1}$, $\eta$ can be 
$5.00 \times 10^{-2}$ to $9.38 \times 10^2$.
\begin{figure}[t!]
  \begin{center}
    \hspace*{-0.5cm} \includegraphics*[width=8cm]{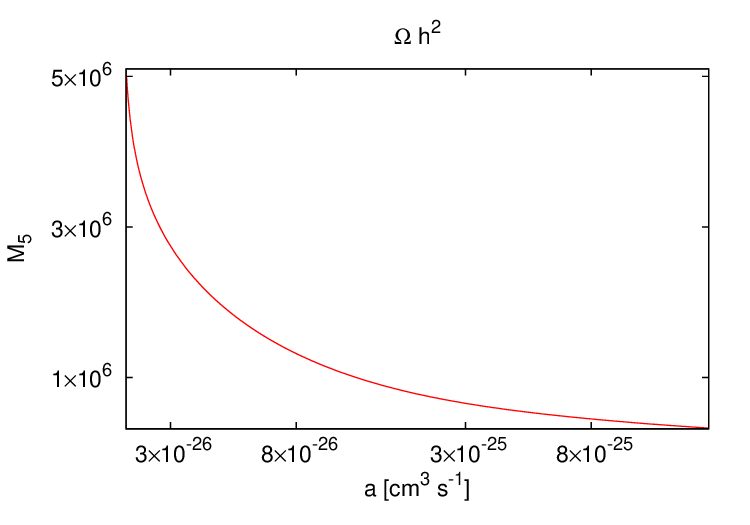}
    \put(-115,-12){(a)}
    \hspace*{-0.5cm} \includegraphics*[width=8cm]{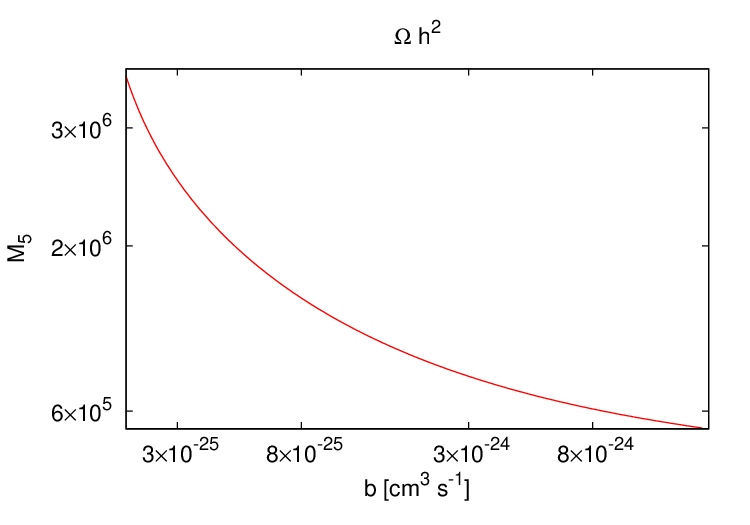}
    \put(-115,-12){(b)}
    \caption{\label{fig:b} 
    \footnotesize Contour plots of the $s-$wave ($p-$wave) annihilation cross 
    section $a$ ($b$) and 
    the 5 dimensional Planck mass $M_5$ when 
    $\Omega_{\rm DM} h^2 = 0.120$ \cite{Planck:2018vyg}.Here  
    $m_{\chi} = 100$ GeV,
    $g_{\chi} = 2$, $g_* = 90$. }
     \end{center}
\end{figure}

In Fig.{\ref{fig:b}}, we obtain the constraints on the 5 dimensional Planck
mass $M_5$ and the cross section of DM in brane world cosmological model when DM 
relic density  
$\Omega_{\rm DM} h^2= 0.120$. $M_5$ can take the values from $3.29 \times 10^5$ GeV to
 $ 5.00 \times 10^6$ GeV for the cross sections 
$2.00 \times 10^{-24} {\rm cm^3} {\rm s}^{-1}$ to 
$2.13 \times 10^{-26} {\rm cm^3} {\rm s}^{-1}$. When $M_5$ is smaller, there
is sizable increased relic
density. Let the DM relic density fall in the range of the observed region,
the cross section must be large. It is just reverse to the case of kination 
model. For the case of $p-$wave annihilation cross sections from
$2.00 \times 10^{-25}{\rm cm^3} {\rm s}^{-1} $ to 
$1.90 \times 10^{-23} {\rm cm^3} {\rm s}^{-1}$, $M_5$ can take 
$3.44 \times 10^6$ GeV to $4.56 \times 10^5$ GeV.

\begin{figure}[h!]
  \begin{center}
     \hspace*{-0.5cm} \includegraphics*[width=8cm]{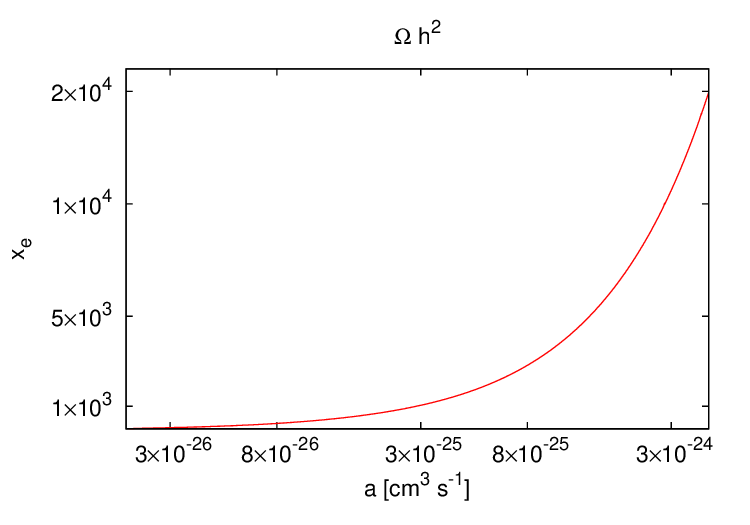}
    \put(-115,-12){(a)}
    \hspace*{-0.5cm} \includegraphics*[width=8cm]{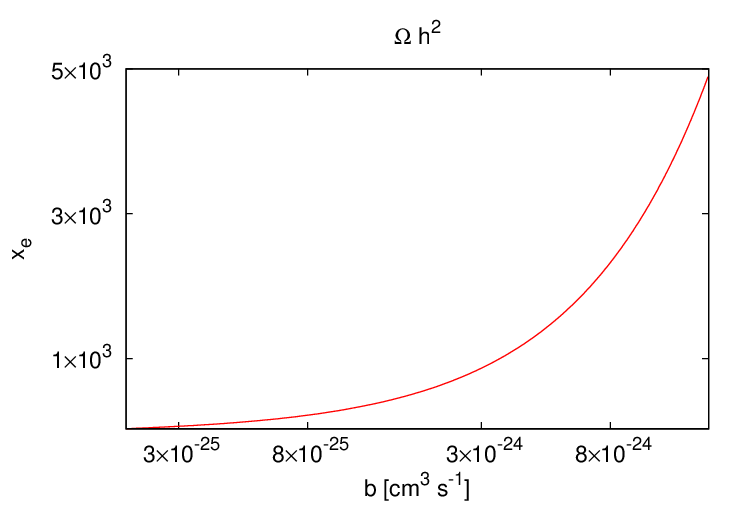}
    \put(-115,-12){(b)}
      \caption{\label{fig:c} \footnotesize
    Contour plots of the $s-$wave ($p-$wave) annihilation cross section 
    $a$ ( $b$) and 
    the inverse-scaled shear temperature $x_e$ for
    $\Omega_{\rm DM} h^2 = 0.120$ \cite{Planck:2018vyg}. Here  
    $m_{\chi} = 100$ GeV,
    $g_{\chi} = 2$, $g_* = 90$. }  
      \end{center}
\end{figure}

Fig.\ref{fig:c} describes the relation between the DM cross section and 
 $x_{\rm e}$ for the values of DM density 
$\Omega_{\rm DM} h^2=0.120$. For the cross sections from 
$2.06 \times 10^{-26} {\rm cm^3} {\rm s}^{-1}$ 
to $4.22 \times 10^{-24}{\rm cm^3} {\rm s}^{-1}$, $x_{\rm e}$ changes from 
$9.54 \times 10^{-2}$ to $1.49 \times 10^4$. For $p-$wave case, when the cross
section changes from $2.02 \times 10^{-25}$ to $1.68 \times 10^{-23}$, $x_{\rm e}$
can be $30.6$ to $4.89 \times 10^3$. There is notable increase of the relic
density for the larger value of $x_{\rm e}$ and we need larger cross section to
obtain the observed value of DM density.   

The parameter spaces which we obtained can be used to test the modified
cosmological models and pave the way of understanding the early universe.

\section{Summary and conclusions}
We review the relic density calculation of DM in the non-standard cosmological
scenarios including kination model, brane world cosmology and Bianchi type I
model. Next, we use the Planck data to constrain the parameter spaces as the
DM cross section and enhancement factor $\eta$ in kination model, 5 dimensional 
Planck 
mass $M_5$ in brane world cosmology and the inverse scaled shear temperature 
$x_{\rm e}$ in shear dominated
universe. The relic density of DM is increased due to the enhanced Hubble
expansion rate in the models which we discussed. The size of the increase of
DM relic density depends on the modification factor. When the modification is
significant, there is sizable increase of the relic density. The cross section 
must be large in order to obtain the observed range of DM relic density. The constraint relationships which are consistent with DM relic density observations are shown in Fig.\ref{fig:a}, Fig.\ref{fig:b} and Fig.\ref{fig:c}.     

Our work is important for the future DM experiments. If we know the cross
sections from DM detections and collider experiments, then we can deduce the
range of modification parameters of the non-standard cosmological models. This is helpful to understand the physics of the very early universe. 

\section*{Acknowledgments}

HI and FL are supported by the National Natural Science Foundation of China
(11765021).

\end{document}